\documentclass{IEEEtran}
\usepackage{cite}
\usepackage{amsmath,amssymb,amsfonts}
\usepackage{graphicx}
\graphicspath{{figures/}}
\usepackage{textcomp,nicefrac}
\usepackage{url}
\usepackage{xcolor}

\def\BibTeX{{\rm B\kern-.05em{\sc i\kern-.025em b}\kern-.08em
T\kern-.1667em\lower.7ex\hbox{E}\kern-.125emX}}
\begin{document}
\title{Real-Time FPGA-Based SiPM Detector Emulation \\ Using a Temporally Quantized Model}

\author{Stefano~Carsi,
        Edoardo~Proserpio,
        Andrea~Abba,
        Francesco~Caponio,
        and~Valentina~Arosio
\thanks{Manuscript submitted June 29, 2026; revised November 27, 2026. This work was carried out at
Nuclear Instruments SRL.}
\thanks{S. Carsi and E. Proserpio are with Nuclear Instruments SRL,
22045 Lambrugo (CO), Italy, and also with the Department of Science and
High Technology, Universit\`a degli Studi dell'Insubria, 22100 Como,
Italy (e-mail: s.carsi@nuclearinstruments.eu;
e.proserpio@nuclearinstruments.eu).}
\thanks{A. Abba, F. Caponio, and V. Arosio are with Nuclear Instruments
SRL, 22045 Lambrugo (CO), Italy.}
}

\maketitle

\begin{abstract}
We present a real-time emulator of silicon photomultiplier (SiPM) signals. Photon-level hits from the open-source SimSiPM
library, which models crosstalk,
afterpulsing, dark counts, and cell recovery, are generated live on the
processor of the system-on-chip or streamed from a host, quantized
upstream into $0.8$~ns bins with the amplitudes summed per bin. A
field-programmable gate array (FPGA) synthesizes the waveform: a
pre-convolution stage preserves sub-cycle timing, and three one-pole recursive filters, look-ahead
transformed to run at $156.25$~MHz, produce the rise and two decays at $1.25$~gigasamples per second (GS/s), interpolated to $2.5$~GS/s. The C simulation of the core, equal
word for word to its register-transfer-level simulation on nine sequences, agrees with a double-precision model of the
binned response to within $1.1$ least-significant bits on the computed
samples, for six sets of time constants and closely spaced pile-up, and
clips at full scale. At the analog output, measured pulses agree with the simulated
core output through a model of the output chain, fitted on one
parameter set and verified on five others, to $0.5\,\%$ RMS of the
peak on the pulse body and $2.4\,\%$ at the peak; the period jitter
is $29$~ps RMS; and one channel plays $55$~million non-empty
bins per second in real time, $78$~million without loss. Live SimSiPM simulations reach the
output with their mean photoelectron number and crosstalk unchanged; calibrated on a real Hamamatsu
S13360 under a pulsed laser, the emulator reproduces the shape of the
sensor's light spectrum within the statistical errors and its
single-photoelectron pulse within $1.8\,\%$ RMS.
\end{abstract}


\section{Introduction}
\label{sec:introduction}
\IEEEPARstart{A}{detector} emulator is an instrument that generates
analog signals reproducing those of a real particle
detector~\cite{caen}. It allows the characterization, debugging, and validation of
front-end electronics---preamplifiers, shapers, analog-to-digital
converters, and trigger logic---without a physical detector or a
radiation source. Because the energy, timing, and multiplicity of every
generated event are known exactly, the emulator also provides perfect
ground truth for the validation of reconstruction algorithms, and it
enables hardware-in-the-loop testing of firmware and data-acquisition
chains under fully reproducible conditions, including controlled rate,
pile-up, and background.

\begin{figure}[t]
\centerline{\includegraphics{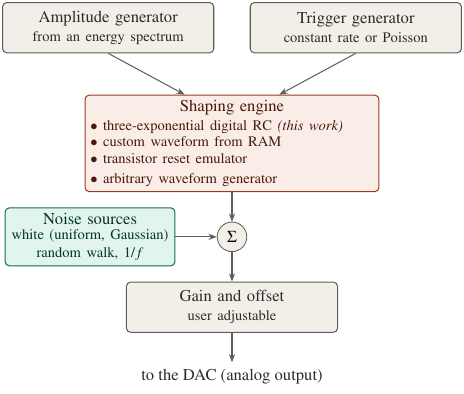}}
\caption{Functional architecture of the detector emulator. An amplitude
generator and a trigger generator drive a configurable shaping engine;
noise, gain, and offset are applied before the digital-to-analog
converter. The three-exponential digital RC shaping core described in
this work (Section~\ref{sec:core}) is one of the available shaping
models.}
\label{fig:concept}
\end{figure}

The key difference between a detector emulator and a conventional
arbitrary waveform generator (AWG) is that the emulator reproduces the
\emph{physics} of the detector rather than replaying a fixed,
precomputed waveform. As sketched in Fig.~\ref{fig:concept}, an amplitude
generator (constant or sampled from an energy spectrum) and a trigger
generator (periodic or Poisson distributed) feed a shaping engine that
synthesizes the detector response; noise, gain, and offset are then
applied before the signal is sent to the output. The shape produced by
the shaping engine, and the parameters that control it, can be changed
at runtime, so that the response of many different detectors can be
reproduced by a single instrument.

Among the detectors that a modern emulator must reproduce, the silicon
photomultiplier (SiPM) is one of the most demanding, owing to its fast
sub-nanosecond rise and its rich microphysics. A natural source of
realistic SiPM events is SimSiPM~\cite{simsipm,proserpio}, an open-source
framework that simulates the SiPM response cell by cell, modeling photon
detection efficiency, optical crosstalk, afterpulsing, dark counts, and
cell recovery at the microscopic level. SimSiPM produces a list of
photon hits, each with a timestamp and an amplitude, at arbitrary
temporal and amplitude resolution; from this list a waveform can be
reconstructed once the detector time constants are fixed. SimSiPM,
however, is a software library: it produces hits, not an analog signal,
and computing the waveform in software at gigasamples per second, in
real time, is out of reach.

Detector emulators based on field-programmable gate arrays (FPGAs) are an
established class of
instruments, both commercial~\cite{caen,caen5810} and in the research
literature~\cite{abba_rt2012,abba_rsi,garcia}, including earlier
instruments developed by some of the present authors~\cite{abba_rt2012,abba_rsi}.
They synthesize the output event by event: an amplitude, drawn from a
programmable energy spectrum, and a trigger time, drawn from programmable
(typically Poisson) statistics, are applied to a pulse whose shape is set
by a few parameters or by a table, and noise and baseline fluctuations
are added digitally. In this event-level paradigm each emulated event is
a single pulse with a single shape. A SiPM signal, however, is the
superposition of many single-cell responses, produced by primary
photoelectrons, crosstalk avalanches, afterpulses, and dark counts, whose
relative sub-nanosecond timing determines the rising edge and therefore
the timing performance of the front-end under test. The emulator
described here differs from the event-level approach in three respects.
First, its input is the photon-level hit list of a microscopic
simulation, so that the statistical microphysics of the detector is
carried into the output hit by hit rather than summarized in an amplitude
distribution. Second, each hit is shaped by a programmable
three-time-constant response with $0.8$~ns timing granularity, obtained
from a $156.25$~MHz fabric clock through a sub-phase pre-convolution,
computed at $1.25$ gigasamples per second (GS/s) and interpolated to the $2.5$~GS/s of the
converters. Third, event generation is decoupled from
shaping, so that the same core can be driven in real time by SimSiPM,
by Geant4, or by any other event source.

In this work we bridge the gap between photon-level simulation and
real-time analog output. We describe a real-time SiPM channel for
a next-generation detector emulator in which SimSiPM runs live on the
processor of a system-on-chip and all signal shaping is performed on an
FPGA. The FPGA receives a compact stream
of photon hits and synthesizes the analog waveform itself, with no
precomputed waveforms, while preserving the full three-time-constant
physical response of the SiPM. We validate the digital shaping core bit
by bit and characterize the analog output with an oscilloscope. The
remainder of the
paper is organized as follows. Section~\ref{sec:model} reviews the SiPM
signal model and the form used in the hardware. Section~\ref{sec:arch}
describes the decoupling of event generation from signal shaping and the
temporal quantization scheme. Section~\ref{sec:core} details the hardware
shaping core and the timing-closure strategy.
Section~\ref{sec:implementation} reports the implementation and resource
usage, Section~\ref{sec:results} validates the digital core bit by bit,
Section~\ref{sec:meas} reports measurements of the analog output,
including live SimSiPM simulations and a comparison with a real SiPM,
and Section~\ref{sec:generalization} discusses generalization and use cases.
Conclusions are drawn in Section~\ref{sec:conclusion}.

\section{SiPM Signal Model}
\label{sec:model}
A SiPM is a matrix of single-photon avalanche diodes (microcells)
operated in Geiger mode and connected in parallel, each in series with a
quenching resistor $R_q$. When a photon is detected in a cell, it
triggers an avalanche that is subsequently quenched; the cell delivers a
charge $Q_{\text{cell}} = (C_d + C_q)\,\Delta V$, where $C_d$ is the
junction capacitance, $C_q$ the parasitic capacitance in parallel with
$R_q$, and $\Delta V$ the overvoltage. In the standard equivalent circuit
of the whole device~\cite{corsi,seifert}, the fired cell is connected in
parallel with the $N-1$ quiescent cells, which form a network
$(N-1)\,C_d$ in series with $R_q/(N-1) \parallel (N-1)\,C_q$, and with the
capacitance $C_g$, the sum of the parasitic capacitances $C_p$ between
the substrate and the contact of the quenching resistor of each cell;
the output node is read through the input resistance $R_S$ of the
front-end. Two limiting time constants govern the decay of the current
delivered to $R_S$:
\begin{align}
\tau_{ff} &\sim R_S\,C_\text{eq}, \qquad
C_\text{eq} = C_g + N\,\frac{C_d\,C_q}{C_d + C_q}, \label{eq:tauff}\\
\tau_{fs} &\sim R_q\,(C_d + C_q), \label{eq:taufs}
\end{align}
the discharge of the output node, whose capacitance on that time scale
is $C_\text{eq}$ (the quenching resistors of the quiescent cells being
bypassed by their capacitances $C_q$), and the recharge of the fired
cell. In large devices the two are coupled, and the measured decay can
exceed both limits: for a $3$~mm Hamamatsu SiPM read on $50~\Omega$,
$R_S$ times the high- and low-frequency capacitances of the device ($99$
and $281$~pF) gives $5$ and $14$~ns, against a measured
$18.6$~ns~\cite{seifert}. The rise is set by the avalanche build-up,
with a time constant of tens of picoseconds~\cite{seifert}, and in
practice by the bandwidth and inductance of the readout, which bring it
to $1$--$2$~ns in the same measurements. For this reason the emulator does not compute its
time constants from the circuit: the four parameters
of~\eqref{eq:superposition} below are fitted to the measured pulse of
the sensor to be emulated, and~\eqref{eq:tauff}--\eqref{eq:taufs} only
indicate their origin and order of magnitude. The three constants correspond to
a fast rise $\tau_r$ (sub-nanosecond to a few nanoseconds at the
output of the readout), a fast
decay $\tau_{ff}$ (nanoseconds to tens of nanoseconds, the discharge of
the output capacitance through the readout), and a slow decay $\tau_{fs}$ (tens to
hundreds of nanoseconds, the recharge through $R_q$). The single-cell
response is commonly written in the product form
\begin{equation}
\label{eq:product}
V(t) \propto \left(1 - e^{-t/\tau_r}\right)
\left[\alpha\,e^{-t/\tau_{ff}} + (1-\alpha)\,e^{-t/\tau_{fs}}\right],
\end{equation}
where $\alpha$ is the fast-component fraction. On top of this
deterministic shape, real SiPMs exhibit dark counts (thermally generated
avalanches), optical crosstalk (a photon emitted during one avalanche
firing a neighboring cell), and afterpulsing (a trapped carrier released
after the avalanche, producing a delayed pulse)~\cite{acerbi}. These effects are
reproduced upstream by SimSiPM and reach the hardware as additional
photon hits, so the shaping core needs to implement only the single-cell
response~\eqref{eq:product}.

For the hardware we adopt the equivalent linear-superposition form
\begin{equation}
\label{eq:superposition}
H(t) = (1 - S_f)\,e^{-t/\tau_{ff}} + S_f\,e^{-t/\tau_{fs}}
       - e^{-t/\tau_r},
\end{equation}
where $S_f$ is the slow-component fraction, so that $\alpha = 1 - S_f$.
Writing $D(t) = (1-S_f)\,e^{-t/\tau_{ff}} + S_f\,e^{-t/\tau_{fs}}$ for
the decay term common to both forms, the difference between the product
form~\eqref{eq:product}, taken with unit proportionality constant, and
the superposition~\eqref{eq:superposition} is
\begin{equation}
\label{eq:moderr}
V(t) - H(t) = e^{-t/\tau_r}\,\bigl[1 - D(t)\bigr] \;\geq\; 0 .
\end{equation}
It vanishes at $t = 0$ and for $t \gg \tau_r$ and, to first order in
$t/\tau_{ff}$, reaches its maximum
$\simeq (\tau_r/e)\,\bigl[(1-S_f)/\tau_{ff} + S_f/\tau_{fs}\bigr]$ at
$t \simeq \tau_r$. The two forms are therefore equivalent when
$\tau_r \ll \tau_{ff}$, with a residual that scales as
$\tau_r/\tau_{ff}$. Table~\ref{tab:moderr} quantifies the residual,
evaluated numerically in double precision and normalized to the peak
of~\eqref{eq:product}. For the representative set of this work the maximum
deviation is $0.72\,\%$ of the peak, located on the rising edge, and the
peak amplitude, peak time, and pulse charge differ by $-0.13\,\%$,
$56$~ps, and $-0.03\,\%$, respectively. The residual grows to a few
percent only for very short fast-decay constants
($\tau_{ff} \approx 10$~ns), and to $7.8\,\%$ in the worst case of the
table, $\tau_r/\tau_{ff} = 0.2$. Because the four parameters
of~\eqref{eq:superposition} are programmable, they can be fitted
directly to a measured single-cell response, in which case the
product form serves only as a physical motivation and the residual
of~\eqref{eq:moderr} does not enter the emulation. The decisive
advantage of~\eqref{eq:superposition} is that it is a \emph{sum} of
three independent exponentials.

\begin{figure}[t]
\centerline{\includegraphics[width=3.5in]{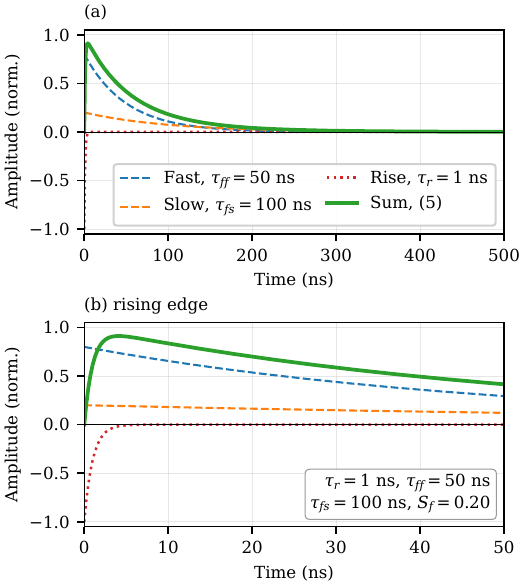}}
\caption{Ideal SiPM response of \eqref{eq:superposition}, shown as the
sum (solid) of its rise, fast-decay, and slow-decay components (dashed),
for typical parameters $\tau_r = 1$~ns, $\tau_{ff} = 50$~ns,
$\tau_{fs} = 100$~ns, and $S_f = 0.20$. (a) $0$--$500$~ns. (b) Zoom
on the rising edge.}
\label{fig:response}
\end{figure}

\begin{table}[t]
\caption{Residual between the product form~\eqref{eq:product} and the
superposition~\eqref{eq:superposition}, as a percentage of the peak
of~\eqref{eq:product}. RMS over $0$--$5\tau_{fs}$; $\Delta Q$: relative
difference of the pulse integrals.}
\label{tab:moderr}
\centering
\setlength{\tabcolsep}{3.5pt}
\begin{tabular}{cccc|cccc}
\hline
$\tau_r$ & $\tau_{ff}$ & $\tau_{fs}$ & $S_f$ & Max. & RMS & Peak & $\Delta Q$ \\
(ns) & (ns) & (ns) & & (\%) & (\%) & (\%) & (\%) \\
\hline
1.0 & 50 & 100 & 0.20 & 0.72 & 0.04 & $-0.13$ & $-0.03$ \\
0.5 & 50 & 100 & 0.20 & 0.35 & 0.01 & $-0.04$ & $-0.01$ \\
2.0 & 50 & 100 & 0.20 & 1.52 & 0.13 & $-0.44$ & $-0.12$ \\
1.0 & 20 & 100 & 0.20 & 1.80 & 0.11 & $-0.57$ & $-0.11$ \\
1.0 & 50 & 200 & 0.50 & 0.49 & 0.02 & $-0.07$ & $-0.01$ \\
1.0 & 10 & 60  & 0.30 & 3.44 & 0.26 & $-1.48$ & $-0.28$ \\
2.0 & 10 & 60  & 0.30 & 7.78 & 0.83 & $-4.55$ & $-1.09$ \\
\hline
\end{tabular}
\end{table} As shown in Fig.~\ref{fig:response},
the total shape decomposes naturally into a rise term and two decay
terms, each of which can be realized as a separate first-order recursive
filter. The four parameters $\tau_r$, $\tau_{ff}$, $\tau_{fs}$, and $S_f$
are programmable at runtime.

\section{System Architecture}
\label{sec:arch}
The central design choice is to \emph{decouple event generation from
signal shaping}. The software side---SimSiPM running on the processing
system, a Geant4 simulation~\cite{geant4} on a host, or any other
source---produces the list of events, that is, \emph{what} happens and
\emph{when}. The FPGA side takes those events
and shapes them into the analog waveform, that is, \emph{how} the detector
responds. Events are transmitted as pairs $(\Delta t, A)$, where $\Delta
t$ is the inter-arrival time with respect to the previous event and $A$
is the summed amplitude. This wire format makes the detector-model
parameters reconfigurable at runtime without re-synthesis, lets a
single shaping core serve many physics models, and reduces the required
bandwidth drastically. Streaming the raw output waveform would require
$2.5~\text{GS/s} \times 16~\text{bit} = 40$~Gb/s per channel, four times
the capacity of a 10-gigabit Ethernet (10~GbE) link. With the event format, the bandwidth
scales with the rate of non-empty sub-phase bins instead of the output
sample rate. Each transmitted bin is encoded in one $b = 64$-bit word,
excluding packet headers: $\Delta t$ in units of $3.2$~ns ($30$~bits,
up to $3.4$~s between consecutive bins), the sub-phase within that unit
($2$~bits), the summed amplitude $A$ ($16$~bits, unsigned fraction of
full scale), and the output channel ($1$~bit); the remaining bits are
reserved. The same word is used on both input paths
(Section~\ref{sec:datapath}).
The 20-photoelectron event of Fig.~\ref{fig:quant} occupies $10$
non-empty bins and is therefore encoded in $10\,b = 640$~bits,
whereas its waveform over $500$~ns amounts to
$1250 \times 16 = 20\,000$~bits, a reduction by a factor of
about~$31$. At a sustained rate of $10^6$ photoelectrons per second
the event stream requires at most $b \times 10^6 = 64$~Mb/s,
more than two orders of magnitude less than the raw waveform;
conversely, the 10~GbE link can carry up to
$10~\text{Gb/s}/b \approx 1.6 \times 10^8$ non-empty bins per
second.

\subsection{Temporal Quantization}
\label{sec:quant}
\begin{figure}[t]
\centerline{\includegraphics[width=3.5in]{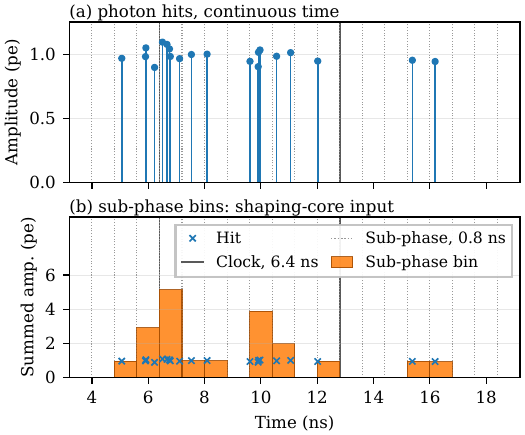}}
\caption{Temporal quantization of a 20-photoelectron event. (a)
Individual photon hits with continuous-time timestamps. (b) The same
hits accumulated into $0.8$~ns sub-phase bins, each spanning
$[kT_S, (k+1)T_S)$, which form the input to the hardware shaping core.
Vertical lines mark the $6.4$~ns coarse clock and the $0.8$~ns
sub-phase grid.}
\label{fig:quant}
\end{figure}

On a processor, photon timestamps have effectively arbitrary precision.
In digital hardware, time is discrete: every operation advances on the
clock edge. The shaping core runs on a fabric clock of $156.25$~MHz,
corresponding to a coarse cycle of $6.4$~ns. To recover sub-cycle timing,
each cycle is subdivided into eight sub-phases of $T_S = 0.8$~ns. The
simulation events are therefore binned upstream into sub-phases. A hit
with timestamp $t_i$ is assigned to the bin
$k_i = \lfloor t_i / T_S \rfloor$, which spans $[k_i T_S, (k_i+1) T_S)$; the amplitudes of all
hits assigned to the same bin are summed, and only the non-empty bins
are transmitted, as the $(\Delta t, A)$ words above, with $\Delta t$
resolved to the sub-phase. No timing
information below $T_S$ is retained: all hits of bin $k$ are shaped as
if they had occurred at $t = k\,T_S$. The amplitudes are summed without
any weighting; the only weighting in the chain is the one applied in
hardware by the pre-convolution stage of Section~\ref{sec:core}, which
propagates each sub-phase amplitude to the following sub-phases with
the decay of the filter. Fig.~\ref{fig:quant} illustrates the binning
for a 20-photoelectron event: the continuous-time hits (a) are
accumulated into the $0.8$~ns sub-phase bins that constitute the
hardware input (b). Pile-up is absorbed natively, because multiple
hits in the same bin simply add.

The value $T_S = 0.8$~ns is therefore the \emph{granularity} of the
trigger timestamps, achieved with a fabric clock of only $156.25$~MHz,
and not the timing accuracy of the output. For a single hit, the floor
operation shifts the pulse earlier by $T_S/2 = 0.4$~ns on average, a
constant offset that can be removed by adding $T_S/2$ to all timestamps
before binning, and adds a random error uniformly distributed over one
bin, with an RMS value of $T_S/\sqrt{12} = 0.23$~ns. For multi-photon
pulses, the errors of the individual hits partially average out. A Monte
Carlo simulation of scintillation-like events (exponential photon arrival
with a $5$~ns time constant, as in the stimuli of Sections~\ref{sec:results}
and~\ref{sec:meas}, the response of Fig.~\ref{fig:response},
and a $20\,\%$ constant-fraction time pick-off) gives an RMS timing error,
after removal of the constant offset, of $0.23$, $0.11$, and $0.05$~ns
for $1$, $20$, and $100$ photoelectrons, respectively. The timing
accuracy of the analog output further depends on the jitter of the
converter clock and output stage, at most $21$~ps RMS per pulse edge
($14$~ps without the deterministic part, Section~\ref{sec:meas-train}),
from two to ten times below these values.
The quantization error per hit is comparable with the single-photon time
resolution of SiPMs, of order $100$~ps: studies that depend on it are
beyond the present scheme, which targets the pulse shape, the charge,
and the timing of multi-photon pulses.

\subsection{Data Path}
\label{sec:datapath}
\begin{figure}[t]
\centerline{\includegraphics{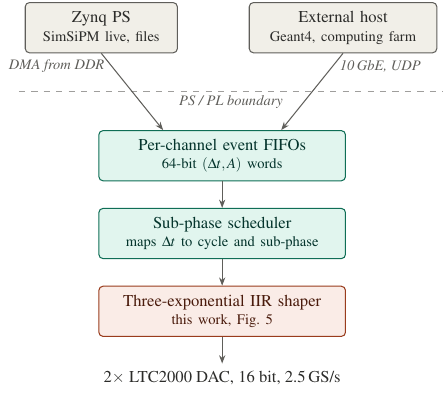}}
\caption{System architecture and data path. Quantized $(\Delta t, A)$
events, one 64-bit word per non-empty sub-phase bin, arrive either from
the processing system (SimSiPM running live, or a file) through a DMA
engine, or from an external host over 10-gigabit Ethernet directly into
the programmable logic. Both paths share the per-channel event FIFOs,
the sub-phase scheduler, and the three-exponential shaping core, which
drives two digital-to-analog converters.}
\label{fig:arch}
\end{figure}

Fig.~\ref{fig:arch} shows the system data path. Quantized events enter
the programmable logic (PL) through one of two paths that carry the same
64-bit word. In the first, the processing system (PS) of the
system-on-chip writes the words into a $64$~MB buffer in its DDR memory,
used as two halves in ping-pong, and a direct-memory-access (DMA) engine
in the PL reads them. In the second, an external host streams them over
10~GbE with the User Datagram Protocol (UDP), to a
UDP/IP stack in the PL that bypasses the PS, so that software does not
become a bottleneck at high event rates. A photon-stream block places
each word on the timeline of its channel, holds per-channel
first-in-first-out (FIFO) buffers, and, at each $6.4$~ns cycle, hands to
the shaping core the eight sub-phase triggers and amplitudes of that
cycle, summing with saturation any words that share a sub-phase. The
stream is lossless: the DMA engine reads only when the FIFOs have room,
the 10~GbE path returns its fill level to the sender, and if the source
falls behind, the timeline pauses, and the pause is counted, instead of
playing hits late. One channel reads up to one word every two fabric
cycles, i.e., $78$~M non-empty bins per second, and every bin is played
on its cycle at any rate: from about $55$~M words per second the
timeline pauses briefly, so that the reading keeps ahead of the shaping
core (Section~\ref{sec:meas-pileup}). The temporal quantization itself is performed
upstream, which keeps the programmable logic lean.

SimSiPM runs on the PS, on the four ARM Cortex-A53 cores of the
system-on-chip, in a separate process that simulates only the photon
hits, not the waveform, bins them, and writes the words into the DMA
buffer. For each fired cell SimSiPM produces a hit with a time and an
amplitude; it models the photon detection efficiency, dark counts
(Poisson in time, on random cells), prompt and delayed optical crosstalk
to neighboring cells, fast and slow afterpulses with exponential delays,
the recovery of a cell (a hit on a cell fired $\Delta t$ earlier has its
amplitude reduced by $1 - e^{-\Delta t/\tau_\text{rec}}$), and a Gaussian
gain spread per hit. Since SimSiPM simulates finite windows, the
continuous timeline is cut into $100~\mu$s slices, each simulated over a
window extended by a tail long enough for afterpulses and delayed
crosstalk, keeping the hits whose primary lies in the slice. The only
approximation, the recovery of cells across a slice boundary, changes
the total charge by less than $6 \times 10^{-8}$, as measured by a
benchmark that compares the amplitudes of every hit on the sliced and
on the continuous timeline. One core
simulates a $3$~mm Hamamatsu S13360-3050 SiPM ($3600$ cells,
$500$~kHz dark-count rate) with $10$~kHz scintillation events of $30$
photoelectrons $3.1$ times faster than real time; the two output
channels run in two threads. Higher rates are generated on a host and
streamed over 10~GbE. The time constants of the shaping core are set
independently, from the response of the emulated SiPM.

\section{Hardware Shaping Core}
\label{sec:core}
\begin{figure}[t]
\centerline{\includegraphics{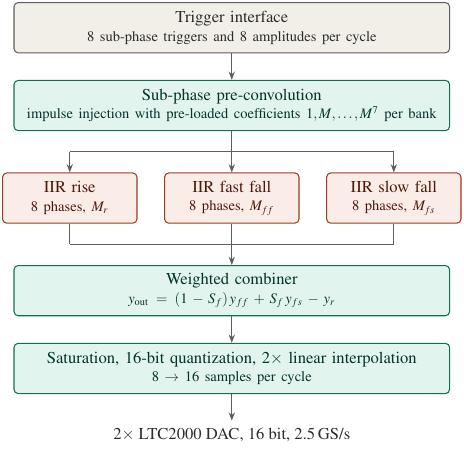}}
\caption{Hardware shaping architecture. The trigger interface delivers
eight sub-phase triggers and eight amplitudes per cycle. A pre-convolution
stage provides sub-phase precision; three parallel one-pole IIR banks
(rise, fast fall, slow fall), each replicated over the eight sub-phases,
are combined and up-sampled before the digital-to-analog converters.}
\label{fig:shaper}
\end{figure}

The shaping core, also referred to as digital RC (DRC) because each of
its filters reproduces digitally the exponential response of a
resistor--capacitor network, is detailed in Fig.~\ref{fig:shaper}. It
exploits the
superposition form~\eqref{eq:superposition}. Each exponential is realized
as a one-pole infinite-impulse-response (IIR) filter
\begin{equation}
\label{eq:iir}
y[n] = M\,y[n-1] + x[n], \qquad M = e^{-T_S/\tau},
\end{equation}
where $T_S = 0.8$~ns is the sub-phase period and $M$ is the pole
corresponding to a given time constant $\tau$. Three such banks run in
parallel---one for the rise, one for the fast decay, and one for the slow
decay---rather than in cascade, which is possible precisely because the
adopted response is a linear superposition. Each bank is replicated
across the eight sub-phases.

\subsection{Sub-Phase Precision via Pre-Convolution}
A trigger that arrives at sub-phase $k$ does not affect only that
sub-phase: it propagates to the later sub-phases of the same cycle and
into the following cycle, weighted by the pole. This sub-phase precision
is obtained with an input pre-convolution stage that injects each impulse
into the eight sub-phase lanes weighted by the pre-loaded coefficients
$1, M, M^2, \ldots, M^7$, a finite-impulse-response kernel of length
eight per bank. The outputs of the three banks are combined by a weighted
combiner that forms
\begin{equation}
\label{eq:combine}
y_\text{out} = (1 - S_f)\,y_{ff} + S_f\,y_{fs} - y_r,
\end{equation}
reproducing~\eqref{eq:superposition}. The result is multiplied by a
programmable gain $G$, which software sets to the inverse of the peak of
the response~\eqref{eq:superposition} (a 16-bit word with $12$
fractional bits), so that an amplitude $A$ produces a pulse of height
$A$ times full scale for any set of time constants. It is then clamped
to the output range, quantized to 16~bits, and up-sampled by a factor of
two through linear interpolation, yielding $16$ samples per cycle at
$0.4$~ns spacing, that is, $2.5$~GS/s. The
filters compute the response at $1.25$~GS/s; the interpolated samples
add no content above $625$~MHz, and their error is quantified in
Section~\ref{sec:results}.

\subsection{Timing Closure: The Look-Ahead Transformation}
\begin{figure}[t]
\centerline{\includegraphics{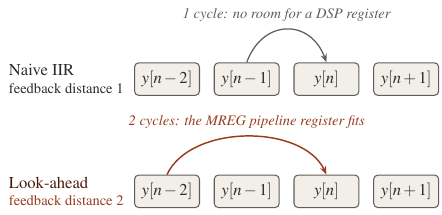}}
\caption{Look-ahead transformation. The naive IIR (top) has feedback
distance one and leaves no room for a pipeline register. The look-ahead
form~\eqref{eq:lookahead} (bottom) has feedback distance two, providing
the two cycles needed to enable the DSP48E2 MREG register while
preserving the same pole and impulse response.}
\label{fig:lookahead}
\end{figure}

The naive recursion~\eqref{eq:iir} has a feedback distance of one: within
a single fabric cycle, the result must be multiplied, accumulated, and
fed back. The internal pipeline registers of the digital signal
processing (DSP) slices cannot be used, because the output is needed in
the same cycle in which it is produced, and the design does not close
timing at $156.25$~MHz. We therefore apply a look-ahead
transformation~\cite{parhi} that rewrites the recursion as
\begin{equation}
\label{eq:lookahead}
y[n] = M^2\,y[n-2] + M\,x[n-1] + x[n],
\end{equation}
which has a feedback distance of two while preserving the same impulse
response and the same pole. As illustrated in Fig.~\ref{fig:lookahead},
two cycles are now available to close the feedback loop, which is exactly
what is needed to enable the MREG pipeline register inside the DSP48E2
slice. Two further measures support timing closure: $M^2$ is precomputed
and latched (read before write), so that no cascaded multiplications occur
in the same cycle; and the inter-phase coefficients are rounded to
27~bits so that every multiplication fits in a single DSP48E2 (a
$27 \times 18$ multiplier). With these techniques the core reaches an
initiation interval of one ($\text{II} = 1$) with three banks
$\times$ eight phases at $156.25$~MHz, meaning that it accepts eight
sub-phase triggers and eight amplitudes on every fabric cycle.

\subsection{Finite-Precision Effects}
\label{sec:precision}
In exact arithmetic,~\eqref{eq:lookahead} has the same impulse response
as~\eqref{eq:iir}: the additional pole at $-M$ introduced by the
look-ahead is cancelled by a zero at the same location. In the
implementation each of the eight lanes of a bank advances by one fabric
cycle, so that the coefficients of~\eqref{eq:lookahead} are the
per-cycle pole $M_c = e^{-8T_S/\tau}$ and its square. Both are unsigned
$16$-bit fractions: $M_c$ is rounded to $16$~bits once by the
software, and $M_c^2$ is computed in hardware as the product of the quantized $M_c$ with itself, rounded to
$16$~bits. The inter-phase coefficients
$e^{-kT_S/\tau}$, $k = 1, \ldots, 7$, have $27$ fractional bits; the
inputs of the banks, the pre-combined inputs, and the state of the
recursion have $24$ fractional bits, i.e., $9$~bits below the output
least-significant bit (LSB); $S_f$ and $1 - S_f$ are $16$-bit
fractions. The integer bits are sized for the range: the inputs of a
lane sum at most eight impulses below unity and have three integer bits,
the pre-combined inputs four, and the state saturates at $8$, so that
the arithmetic is exact as long as each bank stays below eight times the
unit amplitude, i.e., about eight full-scale pulses accumulated in one
bank, and beyond that the output clips instead of wrapping around. In
terms of the input, under a stationary train of rate $R$ and mean
amplitude $\bar A$ (in units of full scale) a bank holds on average
$R \bar A \tau$ with its own time constant, so that the slow bank
reaches its range first, at $R \bar A = 8/\tau_{fs}$, i.e., $80$
full-scale pulses per microsecond for $\tau_{fs} = 100$~ns. For the
parameter sets of Table~\ref{tab:params} the mean output is then already
$3.4$ to $5.4$ times full scale, and the banks saturate only where the
output is clipped; with a small slow fraction and
$\tau_{fs} \gg \tau_{ff}$, for example $(1, 20, 1000~\text{ns}, 0.05)$,
the limit falls at $0.66$ of full scale, and the rate must be kept below
it. Every
multiplication still has one operand of at most $27$~bits and one of at
most $16$~bits (coefficients $M_c$, $M_c^2$, $G$, amplitudes), and fits
in one DSP48E2 slice.

The long-time pole of~\eqref{eq:lookahead} is $\sqrt{M_c^2}$, and the
quantization of the coefficients changes the effective decay time
constant and the area of a bank together, by at most
$3.6 \times 10^{-4}$ up to $\tau = 200$~ns, $1.6 \times 10^{-3}$ up to
$1~\mu$s, and $2.4\,\%$ up to $20~\mu$s; longer constants need
wider coefficients (up to $6\,\%$ between $20$ and $50~\mu$s). The state is rounded to nearest, so that after a
pulse a bank does not decay exactly to zero but stops in a dead band of
$0.5/(1 - M_c^2)$ state LSBs ($2^{-24}$), below $0.1$ output LSB up to
$\tau = 1~\mu$s and $0.8$~LSB at $10~\mu$s. The rise
banks have small poles ($M_c = 0.04$ for $\tau_r = 2$~ns), whose
$16$-bit quantization changes $\tau_r$ by $1.6 \times 10^{-4}$, on a
component that vanishes within a few cycles. For $\tau_r = 1$~ns the
quantized $M_c^2$ is zero, so the rise lane stops after two cycles,
where the exact response has fallen to $e^{-12.8} = 2.8 \times 10^{-6}$
of its initial value, i.e., $0.09$~LSB at full scale; for
$\tau_r = 0.5$~ns $M_c$ itself is zero and the lane stops after one
cycle, at the same residual. The recursion is
stable for any quantized $0 \le M_c^2 < 1$. The effect of all these
quantizations on the waveform was measured directly with the C
simulation of Section~\ref{sec:results}, which runs the synthesizable
source: on the samples computed on the $0.8$~ns grid, the output
differs from the double-precision response of the same binned input by
$-1.1$ to $0$~LSB for all six parameter sets: the truncation of the
output word, between $-1$ and $0$~LSB, plus at most $0.1$~LSB from the
rounding of the amplitudes, coefficients, and state. The output word is
truncated (the conversion of the combiner result to a 16-bit integer,
and the halving in the $2\times$ interpolation, discard the fractional
part), so that the error lies between $-1$ and $0$~LSB, with a
systematic offset of $-0.5$~LSB and an RMS value of $0.29$~LSB about
it; rounding to nearest would remove the offset at the cost of one
adder. Consistently, on the grid samples where the signal is not zero,
the output falls below the reference by $0.51$ to $0.54$~LSB on
average.

\section{Implementation and Resources}
\label{sec:implementation}
\begin{figure}[t]
\centerline{\includegraphics[width=2.6in]{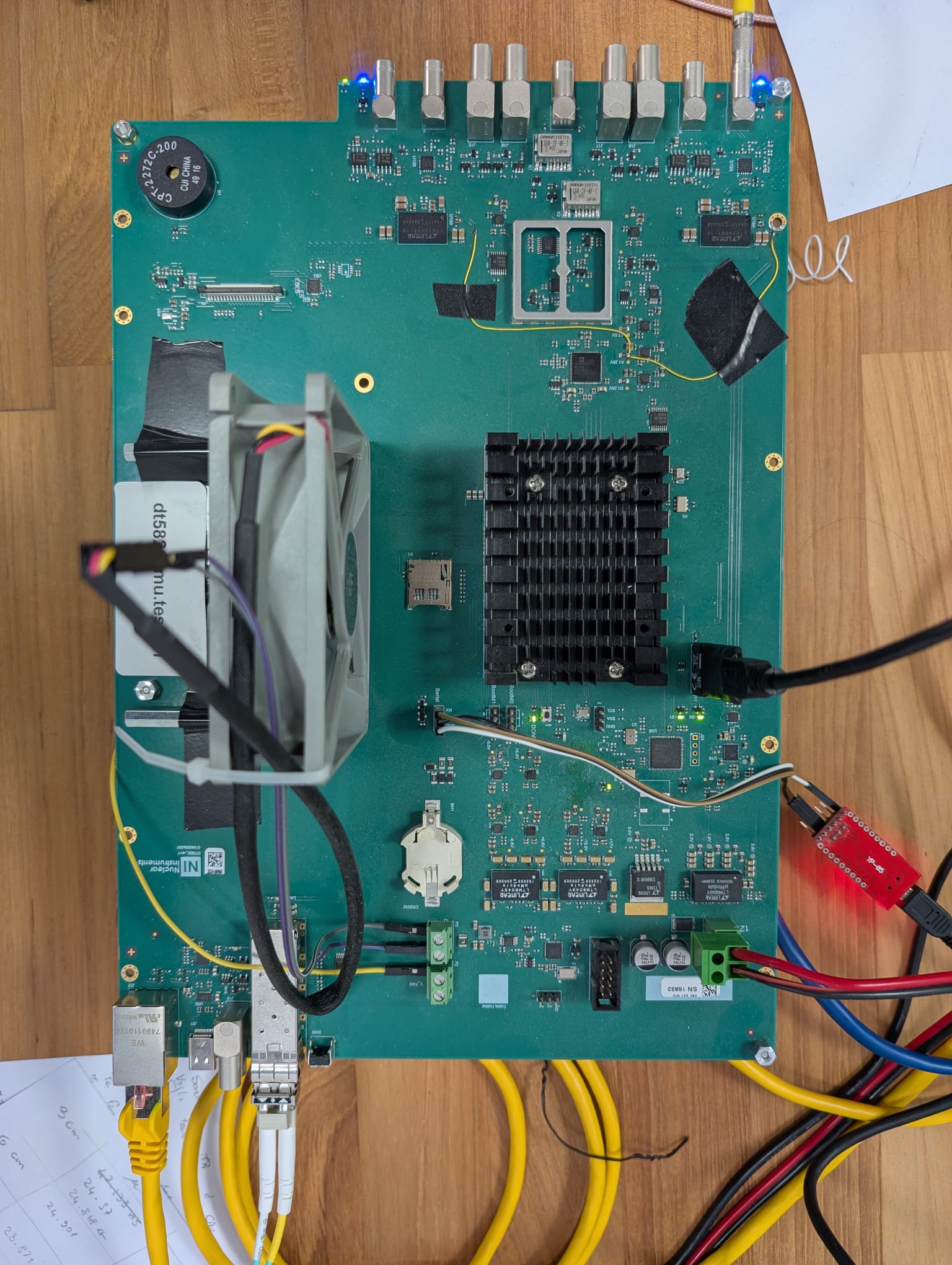}}
\caption{Prototype board. The central heat sink covers the Zynq
UltraScale+ MPSoC; the two LTC2000 digital-to-analog converters and their
analog outputs are at the top left, and the 10-gigabit SFP transceiver
is at the bottom left.}
\label{fig:board}
\end{figure}

The prototype is built around a Zynq UltraScale+ MPSoC~\cite{zynq}
(XCZU6EG-FFVC900-1-I, speed grade $-1$), which
provides both an ARM processing system and FPGA fabric on a single
device. Figure~\ref{fig:board} shows the board: at its center the Zynq
UltraScale+ device; on one side the 10~GbE small-form-factor-pluggable
(SFP) transceiver for the UDP path; and on the other side the two LTC2000
digital-to-analog converters~\cite{ltc2000} (16~bit, $2.5$~GS/s) driving
the two analog channels.

The shaping core is described in Vitis HLS~\cite{vitishls}, that is, in
C++ compiled into register-transfer-level code. High-level synthesis is
well suited here because the algorithm is arithmetic-intensive: the
recursion and the pre-convolution are expressed at the algorithmic level,
and the timing techniques of Section~\ref{sec:core} reduce to a small set
of synthesis directives. The synthesized core achieves an initiation
interval $\text{II} = 1$ with a shaping latency of $16$ cycles
($102.4$~ns). This
latency refers to the shaping core alone, from the sub-phase triggers
and amplitudes at its input to the 16-bit output samples, and is the
value reported by high-level synthesis. It excludes the transfer of the
events over DMA or 10~GbE, the event FIFOs, the sub-phase scheduler,
the converter interface, and the pipeline of the converters themselves;
the delay from the trigger output of the instrument to the analog
output is measured in Section~\ref{sec:meas-latency}.
Table~\ref{tab:resources} summarizes the resource utilization. One
instance of the core occupies $243$ DSP48E2 slices ($12.3\,\%$ of the
device): $168$ in the pre-convolution ($56$ per bank), $48$ in the
look-ahead recursion of the banks, $16$ in the combiner, $8$ for the
gain, and $3$ for $M_c^2$, as listed by high-level synthesis, together
with $17\,677$
flip-flops and $11\,563$ lookup tables after placement and routing, and
no block or ultra RAM. The complete design, with one core per channel
and the other functions of the instrument, meets timing after routing,
over the slow and fast process corners analyzed by Vivado, with a worst
negative slack of $+0.020$~ns and a worst hold slack of $+0.009$~ns.
The two cores take $486$ of the $954$ DSP slices of the design; the
$1019$ slices left on the device would hold four more cores, although
the board has two converters.

\begin{table}[t]
\caption{Resource utilization on the XCZU6EG-1 after placement and
routing, current firmware, used for the comparison with a real SiPM. Core: one instance of the shaping core. Complete
design: two channels and the other functions of the instrument.}
\label{tab:resources}
\centering
\footnotesize
\setlength{\tabcolsep}{3pt}
\begin{tabular}{lccc}
\hline
Resource & Core & Complete design & Available \\
\hline
DSP48E2     & 243 (12.3\,\%)    & 954 (48.4\,\%)      & 1\,973 \\
Flip-flop   & 17\,677 (4.1\,\%) & 236\,312 (55.1\,\%) & 429\,208 \\
LUT         & 11\,563 (5.4\,\%) & 163\,260 (76.1\,\%) & 214\,604 \\
BRAM tiles  & 0                 & 497 (69.6\,\%)      & 714 \\
\hline
\end{tabular}
\end{table}

\section{Validation of the Digital Core}
\label{sec:results}
This section validates the digital shaping chain bit by bit; the analog
output is measured in Section~\ref{sec:meas}, with the same stimuli and
with live SimSiPM simulations. The stimuli of the following tests are
scintillation-like events:
photon arrival times drawn from an exponential distribution with a
$5$~ns time constant, and single-photoelectron amplitudes with a gain
spread of $5\,\%$ RMS. They are produced by a pseudo-random generator in
the C++ testbench, not by SimSiPM: they contain no crosstalk,
afterpulses, or dark counts, which only add further hits of the same
kind and do not exercise the shaping core differently; SimSiPM drives
the measurements of Section~\ref{sec:meas-sipm}. The hit
times are quantized as in Section~\ref{sec:quant}.

The waveforms labeled ``core output'' in Figs.~\ref{fig:valid}
and~\ref{fig:pileup} are obtained by a bit-accurate C simulation of the
shaping core: the synthesizable Vitis HLS source, compiled against the
arbitrary-precision fixed-point libraries of the HLS tool and called by
a C++ testbench. Because high-level synthesis preserves the fixed-point
arithmetic of the source, these samples are the same 16-bit words that
the synthesized core delivers to the digital-to-analog converters for
the same input. This was checked directly on the event of
Fig.~\ref{fig:valid}: the register-transfer-level design generated by
high-level synthesis, simulated in Vivado, delivers on its $16$ output
ports over $200$ cycles the same $3200$ words as the C simulation,
without exception, and so it does on eight further sequences run back to
back, with four sets of time constants changed between them, pile-up,
closely spaced pairs, and events of $1.3$ and $1.6$ times full scale
($27\,840$ words in all). The output was compared with two references: the
analytical response~\eqref{eq:superposition} summed over the binned hits,
each bin placed at the start $kT_S$ of its sub-phase (``binned
reference''), which tests the fixed-point implementation; and the same
sum over the original, unbinned hit times (``ideal''), which shows the
effect of temporal quantization. Both use the amplitudes and the gain
$G$ as quantized at the inputs of the core, and are evaluated at the
$0.4$~ns output spacing in double precision (IEEE~754 binary64), and
stored with at least six significant digits, i.e., to better than
$0.05$~LSB, which is negligible here. The C simulation is untimed and
does not model the pipeline of the synthesized circuit: in all tests
its first non-zero output sample appears $32$~ns (five cycles) after
that of the reference. The offset comes from the source itself: the
algorithm holds its intermediate results in five registers in series
(the latched trigger and amplitudes, the pre-convolved input, its
combination with the previous cycle, the recursion, and the output
latch of the interpolator), modeled as static variables, so that in the
untimed simulation, where one call of the function is one cycle, an
impulse reaches the output five calls later. The
references are shifted by this offset to overlay them on the output. The
offset is therefore not a latency and is unrelated to the $16$-cycle
latency of the synthesized core given in
Section~\ref{sec:implementation}. Throughout this
section, ``quantization noise'' denotes the theoretical quantization
noise of the 16-bit output word ($1/\sqrt{12} \approx 0.29$~LSB RMS),
not the measured analog noise of the output stage. Unless stated
otherwise, the tests use a representative parameter set: $\tau_r = 1$~ns,
$\tau_{ff} = 50$~ns, $\tau_{fs} = 100$~ns, and $S_f = 0.20$; additional
parameter sets are reported in Section~\ref{sec:params}.

\subsection{Single Event}
\begin{figure*}[t]
\centerline{\includegraphics[width=6.8in]{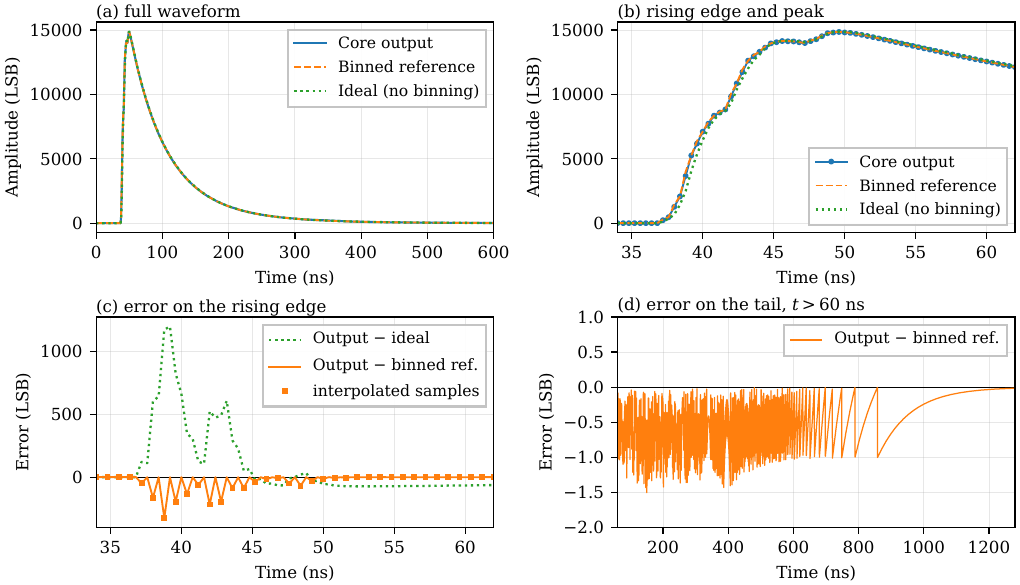}}
\caption{Single-event validation for a 20-photoelectron event. ``Core
output'': 16-bit samples of the shaping core from a bit-accurate C
simulation; the references are shifted onto it by the $32$~ns offset of
the testbench (Section~\ref{sec:results}). (a) Full waveform. (b) Rising
edge and peak; the dots mark the output samples. (c) Error of the core
output with respect to the ideal (unbinned) curve and to the binned
reference on the rising edge; squares: samples produced by the $2\times$
interpolation. (d) Error with respect to the binned reference on the
tail, showing the truncation of the output word.}
\label{fig:valid}
\end{figure*}

Figure~\ref{fig:valid} shows a 20-photoelectron event---the same event
used in Fig.~\ref{fig:quant}. Table~\ref{tab:metrics} reports the
waveform-error metrics: RMS error, maximum absolute deviation, and
peak-amplitude error, in LSB and as a fraction of the pulse peak. The
error is shown in LSB rather than in relative terms, because the
relative error diverges where the signal approaches zero.

The comparison with the binned reference must distinguish the two kinds
of output samples. The samples on the $0.8$~ns grid are computed by the
IIR banks; on these, the output deviates from the binned reference by
$0.52$~LSB RMS and by at most $1.1$~LSB, always by defect, which is the
truncation of the output word discussed in Section~\ref{sec:precision}.
The same behavior produces the saw-tooth pattern on the tail of the
pulse (Fig.~\ref{fig:valid}(d)): as the signal decays, the error ramps
from $0$ to $-1$~LSB between successive output codes. The intermediate
samples are produced by the $2\times$ linear interpolation and deviate
more on the rising edge, by up to $326$~LSB ($2.2\,\%$ of the peak) at
$t = 38.8$~ns (Fig.~\ref{fig:valid}(c)), because a straight line between
two samples $0.8$~ns apart cannot follow a $1$~ns rise. For an isolated
single-cell pulse the interpolation error reaches $6.0\,\%$ of its peak,
$0.4$~ns after the hit, and falls below $2 \times 10^{-4}$ of the peak
after $5\tau_r$. A model made of the binned reference on the $0.8$~ns
grid, followed by $2\times$ linear interpolation and truncation,
reproduces the core output to within one LSB at every sample, and
exactly at $93\,\%$ of the samples: the fixed-point filter behaves as
designed, and the residual is set by the output stage.

With respect to the ideal curve, the deviation is concentrated on the
rising edge, up to $1200$~LSB ($8.1\,\%$ of the peak), where the
$0.8$~ns binning moves hits earlier by $T_S/2$ on average and by up to
one sub-phase on a $1$~ns rise. If the constant offset of $T_S/2$ is
removed (Section~\ref{sec:quant}), the maximum deviation drops to
$537$~LSB ($3.6\,\%$). On the decay the earlier timing leaves a small
negative error of a few tens of LSB, and over the whole pulse the RMS
deviation is $46$~LSB ($0.31\,\%$ of the peak), with a peak-amplitude
error of $-0.19\,\%$.

\begin{table}[t]
\caption{Waveform-error metrics of the core output (C simulation) for
the events of Figs.~\ref{fig:valid} and~\ref{fig:pileup}, from the first
non-zero sample of the reference to the end of the $1.28~\mu$s record.
``$0.8$~ns grid'': only the samples computed by the IIR banks, without
the interpolated ones. Peak error: difference of the highest peaks.}
\label{tab:metrics}
\centering
\setlength{\tabcolsep}{3pt}
\begin{tabular}{lccc}
\hline
 & RMS & Max.\ abs.\ dev. & Peak error \\
 & LSB (\% pk) & LSB (\% pk) & \% \\
\hline
Single, vs.\ binned      & 10.0 (0.07)    & 326 (2.19)  & $0.00$ \\
\quad $0.8$~ns grid      & 0.52 ($<$0.01) & 1.1 (0.01)  & $0.00$ \\
Single, vs.\ ideal       & 46.4 (0.31)    & 1200 (8.06) & $-0.19$ \\
Pile-up, vs.\ binned     & 10.7 (0.09)    & 229 (1.91)  & $-0.06$ \\
\quad $0.8$~ns grid      & 0.57 ($<$0.01) & 1.1 (0.01)  & $0.00$ \\
Pile-up, vs.\ ideal      & 66.2 (0.55)    & 1108 (9.19) & $-0.47$ \\
\hline
\end{tabular}
\end{table}

\subsection{Pile-Up}
\label{sec:pileup}
\begin{figure*}[t]
\centerline{\includegraphics[width=6.8in]{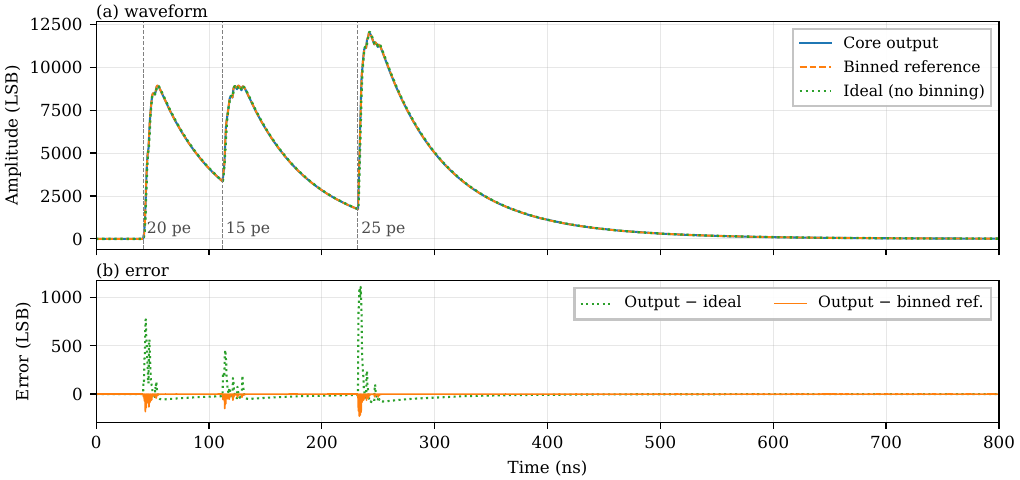}}
\caption{Pile-up validation. Three scintillation-like events of $20$,
$15$, and $25$ photoelectrons, generated at $t = 10$, $80$, and
$200$~ns, overlap on each other's decay tails. ``Core output'' and
alignment as in Fig.~\ref{fig:valid}; the time axis is that of the core
output, so the events (dashed lines) appear $32$~ns later. (a) Waveform.
(b) Error of the core output with respect to the ideal sum and to the
binned reference (metrics in Table~\ref{tab:metrics}).}
\label{fig:pileup}
\end{figure*}

Because the IIR is a linear operator, overlapping events should simply
add. To verify this, three events of $20$, $15$, and $25$ photoelectrons
were generated at $t = 10$, $80$, and $200$~ns, respectively, so that each
event lands on the decay tail of the previous one.
Figure~\ref{fig:pileup} and Table~\ref{tab:metrics} show the same
behavior as for the single event. On the $0.8$~ns grid the core output
stays within $-1.1$ and $0$~LSB of the binned reference ($0.57$~LSB
RMS) over the whole record, including the second and third events, which
start on the decay tails of the previous ones; the largest deviations,
up to $229$~LSB ($1.9\,\%$ of the highest peak), are again those of the
interpolated samples on the three rising edges. With respect to the
ideal curve the deviation reaches $9.2\,\%$ of the highest peak on a
rising edge, and $2.6\,\%$ once the offset of $T_S/2$ is removed. Pile-up is therefore
handled correctly by superposition, without any accumulation of error
in the recursive filters.

Closely spaced hits were tested with two further sets of C simulations,
each pair also simulated with each member alone. Two single-photoelectron
hits separated by $\Delta t = 0.3$~ns fall into the same sub-phase and
are merged into one bin of twice the amplitude; at $\Delta t = 0.8$,
$1.6$, $3.2$, and $6.4$~ns they occupy two bins, in the same cycle or
across the cycle boundary. Two 20-photoelectron events were placed
$0.5$, $1$, and $3$~ns apart, i.e., below and around $\tau_r$. For every
pair, the output of the core for both members together differs from the
sum of the outputs for each member alone by $0$ to $+2$~LSB, which is the
truncation of the three output words and of the summed amplitudes of the
shared bins: superposition holds to the LSB
for any separation, within one sub-phase, within one cycle, and across
cycles. On the $0.8$~ns grid the output stays within $-1.1$ and
$0$~LSB of the binned reference. Below $T_S$, the
temporal information is lost by design: two hits $0.3$~ns apart produce
the same output as one hit of twice the amplitude.

The behavior at full scale was tested with single events of $40$, $60$,
and $80$ photoelectrons at the scale of Fig.~\ref{fig:valid}, whose
binned references peak at $0.89$, $1.30$, and $1.57$ times full scale.
In all three cases the output equals the binned reference clipped at
full scale, to within $\pm 1$~LSB at every sample: the pulse is flat at
full scale while the reference exceeds it and follows the reference
again, without any memory of the overload, as soon as it falls below.
A separate test with eight shapes (including slow ones with a gain $G$
up to $4$), single and piled-up events up to four times full scale,
confirms this, within $2$~LSB plus $3 \times 10^{-4}$ of the reference
(at most $3.3$~LSB) on every grid sample, as long as each bank
stays within its range (Section~\ref{sec:precision}); beyond, under
an extreme overload of about $70$ times full scale, the output stays at
full scale and never collapses.

\subsection{Other Parameter Sets}
\label{sec:params}
The 20-photoelectron event of Fig.~\ref{fig:valid} was simulated again
with five of the other parameter sets of Table~\ref{tab:moderr}: a faster and a
slower rise ($\tau_r = 0.5$ and $2$~ns), a faster fast decay
($\tau_{ff} = 20$~ns), a faster detector ($\tau_{ff} = 10$~ns,
$\tau_{fs} = 60$~ns, $S_f = 0.30$), and a larger and slower slow
component ($\tau_{fs} = 200$~ns, $S_f = 0.50$, $2.56~\mu$s record). Table~\ref{tab:params}
reports the same metrics as Table~\ref{tab:metrics}. In all cases the
samples on the $0.8$~ns grid stay within $-1.1$ and $0$~LSB of the
binned reference, and once the interpolation and the truncation are
modeled the output is reproduced to within $\pm 1$~LSB at every sample
and exactly at $93$ to $95\,\%$ of them. The same event is measured at
the analog output with all six sets in Section~\ref{sec:meas-shape}. The deviation of the interpolated
samples scales with the steepness of the rising edge: it reaches
$5.2\,\%$ of the peak for $\tau_r = 0.5$~ns and drops to $0.78\,\%$
for $\tau_r = 2$~ns.

\begin{table}[t]
\caption{Metrics of the core output (C simulation) for the event of
Fig.~\ref{fig:valid} with six of the parameter sets of
Table~\ref{tab:moderr}. Grid: samples on the $0.8$~ns grid vs.\ the
binned reference. All: all samples vs.\ the binned reference. Ideal:
vs.\ the unbinned hits with the $T_S/2$ offset removed.}
\label{tab:params}
\centering
\footnotesize
\setlength{\tabcolsep}{2pt}
\begin{tabular}{lcccc}
\hline
$(\tau_r, \tau_{ff}, \tau_{fs}, S_f)$ & Peak & Grid & All, max & Ideal, RMS / max \\
ns, ns, ns, -- & LSB & LSB & LSB (\% pk) & \% pk \\
\hline
$(1, 50, 100, 0.20)$   & 14\,854 & $[-1.1, 0.0]$ & 326 (2.19) & 0.16 / 3.6 \\
$(0.5, 50, 100, 0.20)$ & 14\,535 & $[-1.1, 0.0]$ & 758 (5.21) & 0.24 / 5.8 \\
$(2, 50, 100, 0.20)$   & 15\,353 & $[-1.1, 0.0]$ & 119 (0.78) & 0.12 / 2.4 \\
$(1, 20, 100, 0.20)$   & 13\,555 & $[-1.0, 0.0]$ & 355 (2.62) & 0.18 / 4.1 \\
$(1, 10, 60, 0.30)$    & 12\,888 & $[-1.1, 0.0]$ & 389 (3.02) & 0.20 / 4.5 \\
$(1, 50, 200, 0.50)$   & 15\,280 & $[-1.1, 0.0]$ & 319 (2.08) & 0.11 / 3.5 \\
\hline
\end{tabular}
\end{table}

\subsection{Uncertainty Discussion}
The sources of deviation of the digital output from the model are
bounded as follows. (i) The three-exponential form differs from the
product form by $0.72\,\%$ of the peak for the representative set and
by up to $3.4\,\%$ for the six sets used (Table~\ref{tab:moderr}); this term vanishes if the parameters
are fitted directly in the form~\eqref{eq:superposition}. (ii) The
temporal quantization displaces each hit by less than one sub-phase: a
constant offset of $T_S/2$ plus a random error of $0.23$~ns RMS per
hit, reduced to $0.11$~ns RMS on the time pick-off of a
20-photoelectron pulse (Section~\ref{sec:quant}); its effect on the
shape is confined to the rising edge. (iii) The finite-precision
arithmetic changes the decay time constants and DC gains by at most
$3.6 \times 10^{-4}$ in relative terms up to $200$~ns, and its effect on the waveform
is not visible beyond the output truncation: on the $0.8$~ns grid the
output stays within $-1.1$ and $0$~LSB of the double-precision
response for all parameter sets (Section~\ref{sec:precision}). (iv) The truncation of the
16-bit output word contributes an error between $-1$ and $0$~LSB, i.e.,
an offset of $-0.5$~LSB and $0.29$~LSB RMS about it. (v) The $2\times$
linear interpolation introduces a deterministic error on the
intermediate samples, largest where the curvature of the pulse is
largest, i.e., on the rising edge: up to $6.0\,\%$ of the peak for an
isolated single-cell pulse and $2.2\,\%$ for the 20-photoelectron event,
and below $2 \times 10^{-4}$ of the peak after $5\tau_r$. (vi) Above
full scale the output is clipped exactly and recovers without memory,
as long as each bank stays within its range of about eight full-scale
pulses (Section~\ref{sec:precision}). The digital
output therefore reproduces the model to within $1.1$~LSB on the samples
computed on the $0.8$~ns grid; on the rising edge, the temporal
quantization and the interpolation dominate. The bandwidth, noise, and
timing of the analog output stage, which are not included in this
budget, are measured in Section~\ref{sec:meas}.

\section{Measurements of the Analog Output}
\label{sec:meas}
The analog output was measured with a LeCroy HDO4054 oscilloscope
($500$~MHz bandwidth, $12$~bit, $2.5$~GS/s), connected to the output of
channel~0, and where stated also of channel~1, with DC coupling on
$50~\Omega$; the acquisition was driven
remotely, and the raw traces are available from the authors. The
chain of the channel was reduced to the shaping core: unity gain and no
offset in the output stage, and the noise generators of the instrument
(Fig.~\ref{fig:concept}: four sources with a 16-bit amplitude each,
white noise with a uniform or a Gaussian distribution, a random walk, and
$1/f$ noise synthesized by ten first-order sections whose poles are
spaced geometrically between two programmable corners, e.g., $0.5$ and
$7$~MHz) switched off, since the noise that matters for a SiPM, its dark
counts, comes from SimSiPM; their characterization is outside the
scope of this paper. Except for the rate scan of
Section~\ref{sec:meas-pileup} and the measurements of
Sections~\ref{sec:meas-sipm} and~\ref{sec:meas-real}, the stimuli are hit lists like those of
Section~\ref{sec:results}, encoded as 64-bit words and played by the DMA
engine in a loop, with one event every $5~\mu$s. Four builds of the
firmware were used. Most measurements were taken with the first, whose
shaping core differs from that of the other three only by the range of its
internal registers (Section~\ref{sec:precision}) and therefore gives the
same output below full scale: the event of Fig.~\ref{fig:valid},
measured again with the second build, gives the same residual
($0.20\,\%$ RMS). The full-scale events and the trigger-to-output delay
were measured with the second build; the event rates, the bursts, the
10~GbE stream, and the second channel with the third, which has the same
shaping core and differs in the photon-stream block and in functions
outside the core; and the comparison with a real SiPM with the fourth,
the current one of Table~\ref{tab:resources}, which differs from the
third only outside the core and the photon-stream block.

\subsection{Pulse Shape}
\label{sec:meas-shape}
\begin{figure*}[t]
\centerline{\includegraphics{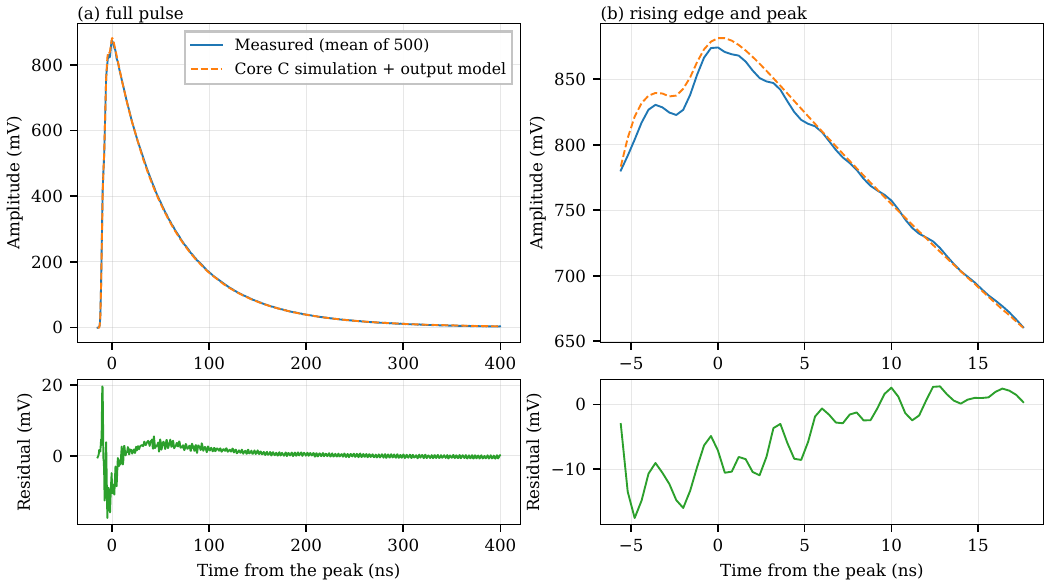}}
\caption{Analog output for the 20-photoelectron event of
Fig.~\ref{fig:valid}, $(\tau_r, \tau_{ff}, \tau_{fs}, S_f) = (1, 50,
100, 0.20)$. Measured: average of $500$ single-shot traces. Model: the
C simulation of the core, through a zero-order hold of $0.4$~ns and the
fitted low-pass model of the output chain. (a) Full pulse. (b) Rising
edge and peak. Bottom: measured minus model.}
\label{fig:scope}
\end{figure*}

Fig.~\ref{fig:scope}(a) compares the average of $500$ single-shot traces
of the event of Fig.~\ref{fig:valid} with the C simulation of the core
for the same input. The digital words are converted with a zero-order
hold of $0.4$~ns, as in the converter, and passed through a model of the
output chain (converter, output network, cable, and oscilloscope): two
real poles, a gain, an offset, and a delay, all fitted on this event.
The fit gives $59.56~\mu$V per LSB, i.e., a full scale of $1.95$~V on
$50~\Omega$, a baseline offset of $+12.0$~mV, and an overall $-3$~dB
bandwidth of $320$~MHz: the fit leaves one significant pole, of
$0.50$~ns, and a second one of $0.01$~ns, so that the model is in effect
a one-pole low-pass. With the
gain and the poles fixed, the event was measured with the other five
parameter sets of Table~\ref{tab:params}; only offset and delay are
fitted, so that these five sets test the model of the output chain on
data not used to build it. The residual
is $0.18$ to $0.21\,\%$ of the peak RMS over the whole pulse (Table~\ref{tab:scope}) and $0.42$
to $0.50\,\%$ on its body ($-15$ to $150$~ns around the peak) for all six
sets. The largest deviations, $1.7$ to $3.0\,\%$ of the peak, occur on
the rising edge, where the response of the output chain is not exactly
that of the low-pass model, and the measured peaks lie $0.3$ to $2.4\,\%$ below the model,
more for faster decays. The rest of the residual has two components. A
fast one, $0.08$ to $0.14\,\%$ of the peak RMS, whose power lies for $76$
to $95\,\%$ at $312.5$~MHz in five of the six sets: the line of
Section~\ref{sec:meas-train}, which does not cancel completely in the
average. A slow one, visible after a $10$~ns moving average: a swelling
of $+0.4\,\%$ of the peak $38$ to $64$~ns after it in all six sets, a
response of the output network that the low-pass model does not
describe. The charge of the pulse, integrated over the window, agrees
with the model to within $0.17\,\%$ for all six sets. Without the
analog low-pass, i.e., comparing the measurement with the zero-order hold
of the digital words alone, the residual grows to $8.9\,\%$ RMS: the
bandwidth of the output chain, not the digital core, shapes the rising
edge at the connector (the $10$--$90\,\%$ rise time of this event is
$5.96$~ns in the digital words and $6.14$~ns at the output). The fitted
$320$~MHz includes the $500$~MHz oscilloscope; for two Gaussian-like
responses the output of the instrument alone would have about
$420$~MHz. The step response of the fitted chain has a $10$--$90\,\%$
rise time of $1.1$~ns with the oscilloscope, about $0.84$~ns without:
rise times below about $1$~ns are reproduced in the digital words but
smoothed at the connector. The
single-shot pulse height varies by $0.2$ to $0.5\,\%$ RMS, at the level
of the baseline noise below.

Channel~1, measured on a second input of the oscilloscope with the same
event, follows the simulation through the model of channel~0, unchanged,
to $0.45\,\%$ of the peak RMS on the body, as channel~0 does; refitted,
its gain differs by $0.05\,\%$. Playing the same events, the two outputs
are $3.71$~ns apart, cables and inputs of the oscilloscope included, with
$31$~ps RMS of variation between single acquisitions. With channel~0
playing and channel~1 silent, the average of channel~1 stays within
$1.1$~mV, $0.12\,\%$ of the peak of channel~0 ($-58$~dB).

\begin{table}[t]
\caption{Analog output: average of $300$--$500$ single-shot traces of
the event of Fig.~\ref{fig:valid} vs.\ the C simulation of the core
through the fitted model of the output chain, $-40$ to $900$~ns around
the peak. Peak error: measured minus model. Spread: single-shot RMS of
the pulse height.}
\label{tab:scope}
\centering
\footnotesize
\setlength{\tabcolsep}{2.5pt}
\begin{tabular}{lccccc}
\hline
$(\tau_r, \tau_{ff}, \tau_{fs}, S_f)$ & Peak & RMS & Max & Peak err. & Spread \\
ns, ns, ns, -- & mV & \% pk & \% pk & \% & \% \\
\hline
$(1, 50, 100, 0.20)$   & 874 & 0.20 & 2.25 & $-0.81$ & 0.41 \\
$(0.5, 50, 100, 0.20)$ & 858 & 0.21 & 2.99 & $-0.27$ & 0.22 \\
$(2, 50, 100, 0.20)$   & 904 & 0.18 & 1.66 & $-0.95$ & 0.21 \\
$(1, 20, 100, 0.20)$   & 789 & 0.19 & 2.38 & $-1.52$ & 0.48 \\
$(1, 10, 60, 0.30)$    & 742 & 0.20 & 2.58 & $-2.41$ & 0.21 \\
$(1, 50, 200, 0.50)$   & 899 & 0.21 & 2.21 & $-1.02$ & 0.36 \\
\hline
\end{tabular}
\end{table}

\subsection{Periodic Train: Timing, Baseline, and Spectrum}
\label{sec:meas-train}
To address the stability of the output in continuous operation, twelve
$500~\mu$s records ($1.25$~GS/s) of the same event repeated every
$5~\mu$s were acquired, $1176$ pulses in all. Measured between pulses of
the same record at $20\,\%$ of their height, with linear interpolation
between samples, the period is $5000.014$~ns (the $2.9$~ppm difference
is that of the oscilloscope and board clocks), with an RMS jitter of
$29$~ps and a maximum deviation of $126$~ps. This jitter is mostly
deterministic: consecutive periods alternate by $\pm 22$~ps, the effect
of the $312.5$~MHz line described below, which makes $1562.5$ periods in
$5~\mu$s and therefore shifts consecutive crossings in opposite
directions, by up to $26$~ps at the slope of the edge. Without the
alternation the period jitter is $19$~ps RMS, i.e., $14$~ps per edge,
of which the amplitude noise of the oscilloscope accounts for $10$~ps on
the period; the sequence of periods shows no pattern repeating every
four pulses, the period with which the hit returns to the same sub-phase
of the clock. Since the sampling clock of the oscilloscope is not
separated, these values are upper bounds for the jitter of the
instrument. The pulse height varies by $0.43\,\%$ RMS. The baseline between pulses has an RMS of $3.7$~mV
($62$~LSB); $85\,\%$ of its power is a single line at $312.5$~MHz, the
data clock of the converter interface ($2.5$~GS/s divided by eight).
The line is deterministic, $4.8$~mV in amplitude, i.e., $-52$~dB with
respect to full scale and $5\,\%$ of the single-photoelectron pulse of
Section~\ref{sec:meas-sipm}; front-ends for spectroscopy filter it out,
fast timing front-ends with gigahertz bandwidth see it. Without that
line the RMS is $1.4$~mV over the $625$~MHz band, close to the
$1.26$~mV of the oscilloscope itself at the same setting, measured on an
unconnected input: the broadband noise of the output is not resolved by
this measurement and is at most about $0.6$~mV RMS.

The spectrum of the train, averaged over the records, consists of lines
at the multiples of $200$~kHz. Their amplitudes agree with the spectrum
of the simulated train through the same output model within $0.07$~dB
RMS about a mean offset of $-0.15$~dB up to $100$~MHz (worst
$-0.29$~dB); above $200$~MHz
the lines approach the noise floor and the $312.5$~MHz line. Each line is
resolution-limited: its power falls within $\pm 2$~kHz, one frequency
bin of the $500~\mu$s records on either side, i.e., the close-in phase noise of the pulse train
is below what these records resolve. Between the lines, the floor is
$-132$~dBm/Hz ($57~\text{nV}/\sqrt{\text{Hz}}$) from $1$ to $100$~MHz,
again at the level of the oscilloscope.

\subsection{Pile-Up and Full Scale}
\label{sec:meas-pileup}
The pairs of Section~\ref{sec:pileup} were measured in the same way. For
the pairs of single-photoelectron hits $0.3$ to $6.4$~ns apart and of
20-photoelectron events $0.5$ to $3$~ns apart, the measured average
follows the C simulation of the pair, through the output model of
Section~\ref{sec:meas-shape}, to $0.33$ to $0.42\,\%$ of the peak RMS,
the same agreement as for the single members; the largest deviations,
up to $5.6\,\%$ of the peak for the single-photoelectron pulses, are on
their $1$~ns rising edges, as above. The full-scale events of
Section~\ref{sec:pileup}, with references at $0.89$, $1.30$, and $1.57$
times full scale, also follow the simulation to $0.18$ to $0.26\,\%$ of
the (clipped) peak RMS: the measured pulses are flat at full scale while the
reference exceeds it and return to the simulated decay without memory of
the overload.

The event rate was raised with Poisson trains of 20-photoelectron events
($0.005$ of full scale per photoelectron, about $11$ words per event),
from $0.1$ to $10$~MHz, played in a loop by the DMA engine. At every
rate no word was lost and every bin was played on its cycle. Up to
$5$~MHz, i.e., $55$~M words per second on one channel, the timeline
paused for less than $0.01\,\%$ of the time. Above, the Poisson
fluctuations of the train exceed more and more often the reading rate of
one word every two fabric cycles, and the timeline pauses, waiting for
the data, as designed: for $0.09\,\%$ of the time
at $66$~M words per second, $2.4\,\%$ at $77$~M, $10\,\%$ at $87$~M,
and $28\,\%$ at $108$~M, where the channel plays $78.1$~M words per
second, its limit of one word every two fabric cycles; the emulated
time then runs correspondingly slower than real time. The mean of the
output, which under heavy pile-up measures the summed charge, agrees
with the rate times the charge of one event ($12.6$~mV per MHz), times
the fraction of time the timeline runs, to within $2.6\,\%$ from $1$ to
$10$~MHz; at $0.3$~MHz the $4$~mV mean is $0.22$~mV higher, within the
DC accuracy of the measurement. Over 10~GbE, a
PC sending the same trains with the flow control of the board delivered
up to $31$~M words per second; at higher offered rates the timeline
paused for the data, again with no word lost or played late. The event
of Fig.~\ref{fig:valid} streamed this way gives the same average as
through the DMA engine to $0.17\,\%$ of the peak RMS.

The instantaneous rate can exceed the sustained one. A scintillation
event of $1000$ photoelectrons, half of full scale at the peak,
follows the simulation to $0.34\,\%$ of the peak RMS. Stretches with
every $0.8$~ns bin occupied, eight words per cycle or sixteen times the
sustained rate, are played on their cycles for $200$~ns ($0.90\,\%$
RMS); longer stretches exhaust the queue of about $63$ cycles that the
reading keeps ahead of the shaping core, and the timeline pauses: no hit
is lost or played late, but the waveform is stretched in time and departs
from the simulation by $2.1\,\%$ RMS for $400$~ns and by $26$ to
$31\,\%$ for $800$ and $1600$~ns.

\subsection{Trigger-to-Output Delay}
\label{sec:meas-latency}
The trigger output of the channel, available on a front-panel LEMO
connector, pulses when the triggers of a cycle reach the shaping core.
With a single-cell pulse repeated every $5~\mu$s, the delay from the
rising edge of the trigger output to the $50\,\%$ point of the rising
edge of the analog pulse
takes four values, $189.7$, $191.2$, $192.9$, and $194.5$~ns: the
trigger output is registered on the $6.4$~ns cycle, while the analog
pulse follows the sub-phase of the hit, and a period of $781.25$ cycles
moves the hit by a quarter of a cycle from one event to the next. Within
each value the delays fall in two clusters about one sample ($0.41$~ns) apart,
selected by the phase of the trigger edge with respect to the sampling
clock of the oscilloscope, which the board does not see: the RMS spread
of $0.2$~ns of each value is an artifact of the acquisition, which leaves
an uncertainty of $\pm 0.2$~ns on the delay, while within a cluster the
spread is $12$~ps RMS, an upper bound for the jitter between the two
outputs. The delay from the trigger output to the analog connector is
therefore $189.7 \pm 0.2$~ns for a hit at the start of a cycle. It is a
delay between two connectors of the instrument, not a de-embedded
latency: it includes the shaping core ($102.4$~ns), the output stage,
the converter interface and pipeline, and the difference between the
trigger-output path and the analog path with their cables; the transfer
of the events and the FIFOs, which depend on the source, come before
it.

\subsection{Live SimSiPM}
\label{sec:meas-sipm}
\begin{figure*}[t]
\centerline{\includegraphics{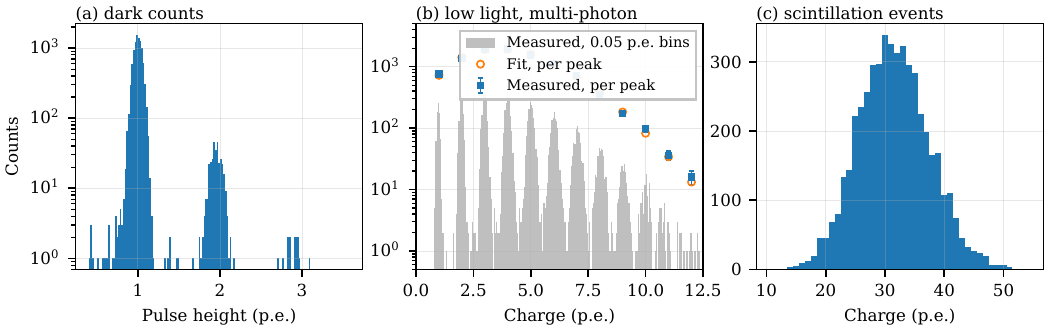}}
\caption{Analog output driven live by SimSiPM with the parameters of a
Hamamatsu S13360-3050 SiPM. (a) Pulse-height spectrum of the dark counts
($10\,000$ pulses). (b) Charge spectrum at low light, dark counts off
($10\,000$ events of $4$ photoelectrons on average): measured counts per
peak compared with the fitted generalized Poisson distribution. (c) Charge spectrum of scintillation
events of $30$ photoelectrons on average ($5000$ events).}
\label{fig:sipm}
\end{figure*}

The emulator was then driven by SimSiPM running live on the
processing system, with the parameters of a Hamamatsu S13360-3050 SiPM
taken from its data sheet ($3$~mm, $50~\mu$m cells, photon detection
efficiency $40\,\%$, dark-count rate $500$~kHz, crosstalk $3\,\%$) and
estimated where the data sheet gives none (afterpulses $1\,\%$ with time
constants of $10$ and $80$~ns, cell recovery $50$~ns, gain spread
$5\,\%$), as an assumption. The shaping core used the generic parameters
of Fig.~\ref{fig:valid}, not those of an S13360 read by a specific
front-end: these measurements test that the statistics produced by
SimSiPM reach the analog output unchanged, not the S13360 model itself,
which is compared with a real sensor in Section~\ref{sec:meas-real}.
In all runs the generator kept pace with real time after a start-up of
about one second, and the output never paused. In a
one-hour run on both channels at once ($10$~kHz of $30$-photoelectron
events and $500$~kHz of dark counts each, $72$ million events and
$5.5 \times 10^{9}$ words in all), the generator
ran at $1.8$ times the real-time rate for the two channels together
(against $3.1$ for the simulation of one channel alone,
Section~\ref{sec:datapath}), and the stream never paused, played no word late, and
lost no word.
Fig.~\ref{fig:sipm} shows three measurements.

With dark counts only and $0.05$ of full scale per photoelectron,
$10\,000$ pulses triggered at half a photoelectron give a single
photoelectron of $96.5$~mV, against $97.6$~mV expected from the fitted
full scale, separated peaks at one, two, and three photoelectrons, and a
fraction of $3.40\,\%$ of pulses of two or more photoelectrons, against
$3.5$ to $3.6\,\%$ expected from a Monte Carlo simulation of the
acquisition and of the analysis: the crosstalk-to-dark-count ratio
counted in the generator, $3.14\,\%$, plus $0.44\,\%$ of accidental
coincidences of two dark counts, within $1.2$ statistical standard
deviations.

With low light and the dark counts switched off, events of a Poisson
number of photons, $10$ on average ($4$ photoelectrons) at $10$~kHz and $0.04$ of full scale per
photoelectron, give the multi-photon charge spectrum of
Fig.~\ref{fig:sipm}(b): peaks from one to eleven photoelectrons, whose
positions lie on a straight line to within $0.08$ photoelectron. A fit of the counts of each peak with a
generalized Poisson distribution with crosstalk~\cite{vinogradov},
truncated at one photoelectron by the trigger, gives $3.994 \pm 0.047$
primary photoelectrons per event, against $4.000$ in the generator, and
a crosstalk probability of $(3.8 \pm 0.9)\,\%$, against $3\,\%$
($\chi^2/\text{ndf} = 16.5/12$); with $10\,000$ events, this test of the
crosstalk is precise to about $1\,\%$ absolute.

With scintillation events of $75$ photons at $10$~kHz (light decay
$40$~ns, $0.01$ of full scale per photoelectron, dark counts on), the
charge integrated over $600$~ns, in units of the charge of one
photoelectron measured in the multi-photon spectrum, is $31.7$
photoelectrons on average with $6.6$ RMS. The generator produces a
Poisson number of photons with $30.0$ primary photoelectrons per event
on average, plus about $0.9$ from crosstalk, $0.3$ from afterpulses, and
$0.3$ from dark counts in the window, i.e., about $31.5$, with a Poisson
width of $5.6$; the integration of the baseline noise over the same
window, measured on the quiet end of the records, adds $2.6$ in
quadrature, for an expected width of about $6.2$ before the excess noise
of crosstalk.

\subsection{Comparison with a Real SiPM}
\label{sec:meas-real}
Finally, the emulator was calibrated on a real sensor with the procedure
of Section~\ref{sec:generalization} and compared with it on the same
oscilloscope. A Hamamatsu S13360-1325 SiPM ($1.3$~mm, $25~\mu$m
cells), operated at about $3$~V of overvoltage at room temperature
(about $25~^\circ$C) and read by a preamplifier of our design, was lit by a
pulsed laser diode (Thorlabs NPL52B, $520$~nm) at $1$~MHz, attenuated to
about $1.5$ photoelectrons per pulse. Its output was acquired like that of the emulator, with DC
coupling on $50~\Omega$ and the trigger at half a photoelectron. The
period of the laser follows from the trigger times, and the triggers are
sorted by their phase: those within $\pm 15$~ns of a laser pulse form
the light spectrum, those from $300$~ns after a pulse to $200$~ns before
the next one the dark counts. These include background light: the
sensor was not in a light-tight enclosure, and light from the indicators
of the instruments around it reached it.

The sensor gives $61.7$~mV per photoelectron. Its average
single-photoelectron pulse ($377$ pulses without a second pulse in the
record), fitted with~\eqref{eq:superposition} through the output chain
of the emulator, gives $\tau_r = 5.95$~ns, $\tau_{ff} = 7.73$~ns,
$\tau_{fs} = 32.0$~ns, and $S_f = 0.025$, with a residual of $1.1\,\%$
RMS of the peak. These are effective constants of the sensor and its
readout, not of the bare sensor: the preamplifier rounds the rising
edge more than the exponential rise of~\eqref{eq:superposition} can
follow, and the largest deviations, up to $13\,\%$ of the peak, are
there. The fit is constrained to $\tau_r < \tau_{ff} < \tau_{fs}$:
without the constraint it converges to two nearly equal time constants
with coefficients of opposite sign, which, normalized to the peak, would
overflow in the core. Between the laser pulses, the sensor has a
dark-count rate of $305 \pm 12$~kHz, counted on free-running records of
$200~\mu$s, and $(1.6 \pm 0.4)\,\%$ of its dark pulses reach
$1.5$~photoelectrons or more. Its single-photoelectron peak is
$5.01\,\%$ RMS wide, of which $2.46\,\%$ is noise, i.e., a gain spread of
$4.36\,\%$. With the laser on, the records of the sensor also contain
laser pulses: $28.7\,\%$ of its single-photoelectron records have a
second pulse, against $12.0\,\%$ for the emulator with dark counts only, so that afterpulses could not
be measured and are not compared. A fit of the light spectrum with a Poisson number
of primaries, each with a geometric number of crosstalk companions,
gives $1.453 \pm 0.016$ primary photoelectrons per pulse.

\begin{figure*}[t]
\centerline{\includegraphics{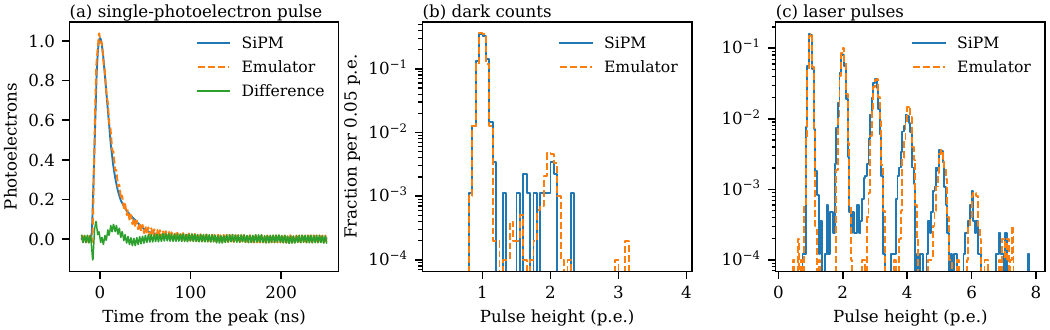}}
\caption{Hamamatsu S13360-1325 SiPM under a pulsed laser and the emulator
calibrated on it, on the same oscilloscope, in photoelectrons. (a)
Average single-photoelectron pulse and difference. (b) Pulse-height
spectrum of the dark counts ($882$ sensor pulses between the laser
pulses, $10\,000$ emulator pulses). (c) Pulse-height spectrum of the
laser pulses ($1.45$ primary photoelectrons on average; $8337$ and
$10\,000$ triggered pulses).}
\label{fig:real}
\end{figure*}

The emulator was then programmed with these values: the four
constants in the shaping core, $0.05$ of full scale per photoelectron,
and in SimSiPM the measured dark-count rate, crosstalk, and gain spread,
the other parameters as in Section~\ref{sec:meas-sipm}. The vertical
scale of the oscilloscope was set in proportion, so that both signals
are analyzed identically, in photoelectrons. Fig.~\ref{fig:real} shows
the comparison. With dark counts only, $(1.87 \pm 0.14)\,\%$ of the
emulator pulses reach $1.5$~photoelectrons ($+0.6$ standard deviations
from the sensor) and the dark-count rate is $306.9 \pm 3.3$~kHz
($+0.2$); since both were programmed from the sensor, these two figures
close the loop through SimSiPM and the output rather than test the
model. Beyond the programmed values, the pulse tests the fit and the
output chain, and the light spectrum tests the shape of the
distribution for the mean fitted on it. The average single-photoelectron pulses differ by $1.8\,\%$
RMS of the peak, by $2.4\,\%$ at the peak, and by $4.7\,\%$ in charge,
and by up to $10\,\%$ on the rising edge; the emulator alone follows
its programmed response within $1.3\,\%$ RMS, more than in
Table~\ref{tab:scope} because the pulse is only $0.05$ of full scale, so
the $312.5$~MHz line of the output weighs correspondingly more. The
difference from the sensor thus comes about equally from the fit
($1.1\,\%$) and from the output ($1.3\,\%$). For the light, SimSiPM
generated events of a Poisson number of primary photoelectrons of mean
$1.453$ (detection efficiency set to one), with the dark counts off, at
$200$~kHz rather than $1$~MHz, because random events at the laser rate
would pile up where the periodic laser does not. The photoelectrons of an
event arrive all at once, while the laser spreads them over its pulse
width, of at least $5$~ns; the agreement of the mean heights shows that
this difference is below the precision of the comparison. The
fractions of pulses from one to seven photoelectrons agree within
$2.0$ standard deviations ($\chi^2 = 9.1$ for $5$ degrees of freedom),
and the mean pulse heights within $1.1\,\%$. Thus, calibrated on a
sensor, the emulator returns the programmed dark-count statistics and
reproduces the shape of the light spectrum of the sensor within the
statistical errors of these measurements and its pulse within
$1.8\,\%$ RMS, except for the rising edge set by the
preamplifier.

\section{Generalization and Use Cases}
\label{sec:generalization}
The primary use case is a front-end testbench: the emulator output is
connected directly to the input of the front-end under development, which
then sees signals that reproduce those of a real SiPM within the limits
measured in Section~\ref{sec:meas}, with perfect ground truth and full reproducibility, and without any detector or
radioactive source. Beyond SiPMs, the same hardware generalizes in
several directions. Coupled with Geant4~\cite{geant4}, particle-level
interactions can be transformed directly into analog detector outputs in
real time. The same core can also emulate the differing pulse shapes of
gamma rays and neutrons in organic scintillators for
pulse-shape-discrimination studies: the firmware has a mode, not
evaluated in this paper, in which each event carries its own
slow-component weight, delivered as separate impulses to the fast and
slow banks, so that events with different shapes pile up exactly. More generally, the shaping core is a
primitive: any detector whose response is described by three time
constants can be emulated without re-synthesis, simply by reprogramming
the four parameters.

\emph{Calibration.} To emulate a given sensor read by a given
front-end, the four parameters of~\eqref{eq:superposition} are fitted to
the average single-photoelectron pulse of the sensor at the output of
its readout. Because the emulator adds its own output chain
(Section~\ref{sec:meas-shape}), the comparison is made between the
sensor pulse and the emulator output, i.e., the core response through
the measured low-pass, not the bare core response. The amplitude per
photoelectron follows from the single-photoelectron height or charge of
the sensor and the measured full scale ($1.95$~V on $50~\Omega$,
$59.6~\mu$V per LSB). The SimSiPM parameters, i.e., detection
efficiency, dark-count rate, crosstalk, afterpulsing, recovery, and gain
spread, come from the data sheet or, better, from the dark spectrum of
the sensor itself, as in Section~\ref{sec:meas-real}.

\emph{Limits.} The emulator is a voltage source behind $50~\Omega$, while
a SiPM is a current source with a capacitance of hundreds of
picofarads (Section~\ref{sec:model}): front-ends whose response depends
on the source impedance, such as charge-sensitive or high-impedance
voltage amplifiers, see a different source unless an adapter network
reproduces it, and the emulator then provides the pulse of the sensor
as seen by the readout. The amplitude of a bare sensor is small: one
photoelectron of an S13360-3050 at a gain of $1.7 \times 10^{6}$ carries
$0.27$~pC, i.e., $0.2$ to $0.7$~mV at the peak on $50~\Omega$ for decay
times of $60$ to $19$~ns, $4$ to $12$~LSB of the converter and below the
$4.8$~mV of the $312.5$~MHz line. For single-photoelectron studies the
emulator is therefore set to a larger photoelectron and followed by an
attenuator: at $0.05$ of full scale per photoelectron the line is
$5\,\%$ of it, at the cost of a range of $20$ photoelectrons. For
multi-photon signals smaller amplitudes are usable, because a charge or
a shaped measurement integrates the line out, as in
Fig.~\ref{fig:sipm}(c) at $0.01$ of full scale per photoelectron. It is
the line, not the resolution of the converter, that sets this limit.

\section{Conclusion}
\label{sec:conclusion}
We have demonstrated a real-time emulation of silicon photomultiplier
signals on an FPGA, from photon-level simulation to the analog output.
SimSiPM runs live on the processing system, or hits are streamed over
10~GbE, and the FPGA synthesizes the full three-time-constant SiPM
response from a compact stream of 64-bit words, producing a 16-bit
output at $2.5$~GS/s with an initiation interval of one. A
temporal-quantization scheme combined with a sub-phase pre-convolution
delivers a trigger timestamp granularity of $0.8$~ns, corresponding to a
timing error of $0.23$~ns RMS per hit, from a $156.25$~MHz fabric clock,
and a look-ahead transformation of the recursive filter makes timing
closure possible without measurable loss of numerical accuracy. A
bit-accurate C simulation of the synthesizable core agrees with a
double-precision model of the binned response to within $1.1$~LSB on
the samples computed by the filters, for six sets of time constants and
closely spaced pile-up, and the output clips at full scale.
At the analog output, the measured pulses agree with the simulated core
output through a model of the $320$~MHz output chain, fitted on one
set and verified on five others, to $0.5\,\%$ RMS of the peak on the
pulse body and $2.4\,\%$ on the peak; the period of a pulse train
jitters by $29$~ps RMS; one channel plays $55$~million words per second
in real time and up to $78$~million without loss, and its output stays linear
under heavy pile-up; and the
multi-photon spectrum of a live SimSiPM simulation of a Hamamatsu S13360
SiPM returns its mean number of photoelectrons within its statistical
error of $1.2\,\%$, and its crosstalk within its statistical error of
$0.9\,\%$ absolute.
Calibrated on a real Hamamatsu S13360-1325 SiPM under a pulsed laser, the
emulator reproduces the shape of the light spectrum of the sensor within
the statistical errors and its average single-photoelectron pulse
within $1.8\,\%$ RMS of the peak, and returns the dark-count rate and crosstalk programmed from it. The architecture is general: the same shaping core serves any
detector with a three-time-constant response, which makes the emulator a
flexible instrument for hardware-in-the-loop testing of front-end
electronics. Future work will pursue comparisons with other sensors and
readouts, including a source network that reproduces the impedance of
the sensor, end-to-end tests with front-end electronics, and a direct
Geant4-to-converter pipeline.

\end{document}